\documentclass[12pt]{article}
\usepackage{amsmath}
\usepackage{amsfonts}
\usepackage{amsthm}
\usepackage{authblk}
\usepackage{cite}
\usepackage{graphicx}
\usepackage[colorlinks=true]{hyperref}
\newtheorem{thm}{Theorem}

\newtheorem{cor}[thm]{Corollary}
\newtheorem{lem}[thm]{Lemma}
\newtheorem{prop}[thm]{Proposition}
\theoremstyle{definition}
\newtheorem{defn}[thm]{Definition}
\theoremstyle{remark}
\newtheorem{rem}[thm]{Remark}

\begin{document}

\title{The Entropy Power Inequality with quantum conditioning}
\author{Giacomo De Palma}
\affil{QMATH, Department of Mathematical Sciences, University of Copenhagen, Universitetsparken 5, 2100 Copenhagen, Denmark}
\date{}
\maketitle

\begin{abstract}
The conditional Entropy Power Inequality is a fundamental inequality in information theory, stating that the conditional entropy of the sum of two conditionally independent vector-valued random variables each with an assigned conditional entropy is minimum when the random variables are Gaussian.
We prove the conditional Entropy Power Inequality in the scenario where the conditioning system is quantum.
The proof is based on the heat semigroup and on a generalization of the Stam inequality in the presence of quantum conditioning.
The Entropy Power Inequality with quantum conditioning will be a key tool of quantum information, with applications in distributed source coding protocols with the assistance of quantum entanglement.
\end{abstract}
\section{Introduction}
The Shannon differential entropy \cite{cover2006elements} of a random variable $X$ with values in $\mathbb{R}^n$ and probability density function $p(x)$ is
\begin{equation}\label{eq:S(X)}
S(X) = -\int_{\mathbb{R}^n}p(x)\ln p(x)\,\mathrm{d}x\;,
\end{equation}
and quantifies the information got when the value of $X$ is revealed.

A fundamental problem in information theory is the following.
Given $a,\,b\in\mathbb{R}$, let $X$ and $Y$ be independent random variables with values in $\mathbb{R}^n$ such that $S(X)=a$ and $S(Y)=b$.
What is the minimum possible value of $S(X+Y)$?
The Entropy Power Inequality \cite{dembo1991information,stam1959some,shannon2001mathematical} states that this minimum is achieved when $X$ and $Y$ have a Gaussian probability distribution with proportional covariance matrices, and reads
\begin{equation}\label{eq:clEPI}
\exp\frac{2S(X+Y)}{n} \ge \exp\frac{2S(X)}{n} + \exp\frac{2S(Y)}{n}\;.
\end{equation}
The Entropy Power Inequality is a fundamental tool of information theory \cite{cover2006elements}.
It was introduced by Shannon to provide an upper bound to the information capacity of non-Gaussian channels \cite{shannon2001mathematical}, and was later employed to bound the information capacity region of the Gaussian broadcast channel \cite{bergmans1974simple} and the secret information capacity of the Gaussian wiretap channel \cite{leung1978gaussian} and to prove the convergence in relative entropy for the central limit theorem \cite{barron1986entropy}.
The Entropy Power Inequality was also extended to the R\'enyi entropies \cite{bobkov2017variants} and to the framework of free probability theory \cite{szarek1996volumes}.

Let the random variable $X$ with values in $\mathbb{R}^n$ be correlated with the random variable $M$ with values in the countable set $\mathcal{M}$.
For any $m\in\mathcal{M}$, let $q(m)$ be the probability that $M=m$, and let $p(x|m)$ be the probability density of $X$ conditioned on $M=m$.
The Shannon differential entropy of $X$ conditioned on $M$ is \cite{cover2006elements}
\begin{equation}\label{eq:defccS}
S(X|M) = \sum_{m\in\mathcal{M}} q(m)\,S(X|M=m)\;,
\end{equation}
where for any $m\in\mathcal{M}$
\begin{equation}
S(X|M=m) = -\int_{\mathbb{R}^n} p(x|m) \ln p(x|m)\,\mathrm{d}x
\end{equation}
is the Shannon differential entropy of $X$ conditioned on $M=m$.

The conditional Entropy Power Inequality \cite{wang2017distributed,de2018conditional} solves the conditional version of the problem above.
Given $a,\,b\in\mathbb{R}$, let $X$ and $Y$ be random variables with values in $\mathbb{R}^n$ correlated with the random variable $M$ with values in $\mathcal{M}$ such that $S(X|M)=a$ and $S(Y|M)=b$.
Let us assume that $X$ and $Y$ are conditionally independent given $M$, i.e., for any $m\in\mathcal{M}$, $X$ and $Y$ are independent if conditioned on $M=m$ (if $\mathcal{M}$ is trivial, the conditional independence reduces to the standard independence).
What is the minimum value of $S(X+Y|M)$?
The conditional Entropy Power Inequality \cite{wang2017distributed,de2018conditional} states that this minimum is achieved when, for any $m\in\mathcal{M}$, $X$ and $Y$ conditioned on $M=m$ have a Gaussian probability distribution with proportional covariance matrices, and the proportionality constant does not depend on $m$.
The inequality is identical to the Entropy Power Inequality \eqref{eq:clEPI} with the entropies replaced by the conditional entropies:
\begin{equation}\label{eq:clcondEPI}
\exp\frac{2S(X+Y|M)}{n} \ge \exp\frac{2S(X|M)}{n} + \exp\frac{2S(Y|M)}{n}\;.
\end{equation}
The conditional Entropy Power Inequality is a key tool in information theory.
Its main application is in the field of distributed source coding \cite{wang2017distributed}, which is a fundamental problem of network information theory concerning the compression of multiple correlated information sources that do not communicate with each other.
In this field, the conditional Entropy Power Inequality is necessary to prove the converse theorems for the quadratic Gaussian CEO problem with two terminals \cite{oohama2005rate,prabhakaran2004rate,wang2017distributed} and for the Gaussian multi-terminal source coding problem with two sources \cite{gamal2011network,wang2017distributed}.
In the quadratic Gaussian CEO problem with two terminals there are two encoders, each of which observes a noisy version of the same signal assumed to have a Gaussian distribution.
The goal of each encoder is to encode his observation using the minimum number of bits such that the original signal can be recovered from the encodings with a given bound on the average error.
In the Gaussian multi-terminal source coding problem with two sources, each encoder observes a perfect copy of a different signal, and the signals observed by the two encoders are correlated.

\subsection{Our contribution}
We prove the conditional Entropy Power Inequality \eqref{eq:clcondEPI} in the scenario where $M$ is a quantum system, $X$ and $Y$ are conditionally independent given $M$ and $S(X+Y|M)$, $S(X|M)$ and $S(Y|M)$ are the quantum conditional entropies (Theorem \ref{thm:EPI}, see \autoref{sec:XM} for the definitions).

A similar result has been proven for the sum of binary random variables \cite{hirche2018bounds}.

Entropic inequalities with classical conditioning easily follow from the corresponding unconditioned inequalities and from the definition \eqref{eq:defccS} of conditional entropy via Jensen's inequality (see e.g. \cite[Appendix A]{de2018conditional} or \cite[Appendix A.5.2]{wang2017distributed}).
Since the quantum conditional entropy cannot be defined as in \eqref{eq:defccS} \cite{beigi2014dimension}, the same method cannot be applied to entropic inequalities with quantum conditioning, whose proof is very challenging.

The Entropy Power Inequality \eqref{eq:clcondEPI} with quantum conditioning complements the conditional Entropy Power Inequality for bosonic quantum systems \cite{koenig2015conditional,de2018conditional} and the conditional Entropy Power Inequality for quantum additive noise channels \cite{huber2018conditional}.
In the conditional Entropy Power Inequality for bosonic quantum systems \cite{koenig2015conditional,de2018conditional}, the random variables $X$ and $Y$ are replaced by quantum Gaussian systems modeling the electromagnetic radiation and the sum $X+Y$ is replaced by the beam-splitter operation \cite{weedbrook2012gaussian,serafini2017quantum}.
In the conditional Entropy Power Inequality for quantum additive noise channels \cite{huber2018conditional}, $X$ is replaced by a quantum Gaussian system and $Y$ is still a random variable representing classical noise added to the quantum state of the radiation.

The Entropy Power Inequality proven in this paper is optimal both with and without quantum conditioning.
Indeed, if the quantum system $M$ is trivial and $X$ and $Y$ are independent Gaussian random variables with proportional covariance matrices, equality is achieved in \eqref{eq:clcondEPI}.
On the contrary, the presence of quantum conditioning is fundamental for the optimality of the Entropy Power Inequalities for quantum Gaussian systems of Refs. \cite{koenig2015conditional,de2018conditional,huber2018conditional}.
If the quantum system $M$ is not present, quantum Gaussian states do not achieve equality, hence the corresponding unconditioned inequalities are not optimal \cite{konig2014entropy,konig2016corrections,de2014generalization,de2015multimode,de2017gaussian,huber2017geometric}.
The optimal inequality that would be saturated by quantum Gaussian states is the longstanding conjecture open since 2007 called Entropy Photon-number Inequality \cite{guha2008entropy}, which has so far been proven only in some particular cases \cite{de2015passive,de2016passive,de2016gaussian,qi2017minimum,de2018pq,de2016gaussiannew,de2017multimode,de2018new} (see \cite{holevo2015gaussian,de2018gaussian} for a review).
The unconditional Entropy Power Inequality has been explored also for other quantum systems and operations, e.g. the partial swap for qudits \cite{audenaert2016entropy}.

Both in the classical and in the quantum scenarios, the prominent proofs of Entropy Power Inequalities are based on perturbation with the heat semigroup and on some version of the Stam inequality for the Fisher information \cite{stam1959some,de2018conditional,koenig2015conditional,huber2018conditional,konig2014entropy,de2014generalization,de2015multimode,de2018new}.
In this paper we follow the same approach: our proof of the Entropy Power Inequality with quantum conditioning is based on the perturbation with the quantum version of the heat semigroup and on a new Stam inequality for the Fisher information with quantum conditioning (Theorem \ref{thm:Stam}).
Other approaches to prove quantum entropic inequalities are majorization theory \cite{giovannetti2015solution,holevo2015gaussian,de2015passive,audenaert2016entropy} and Lagrange multipliers \cite{de2016gaussian,de2018pq}.

The paper is structured as follows.
In \autoref{sec:XM} we define the conditional entropy and the conditional independence of random variables conditioned on a quantum system.
In \autoref{sec:Stam} we prove the Stam inequality with quantum conditioning (Theorem \ref{thm:Stam}), and in \autoref{sec:EPI} we prove the Entropy Power Inequality with quantum conditioning (Theorem \ref{thm:EPI}).
We conclude in \autoref{sec:concl}.

\section{Random variables correlated with a quantum system}\label{sec:XM}
\subsection{Conditional entropy}
A state \cite{holevo2013quantum,wilde2017quantum} of the quantum system $M$ is a positive linear operator with unit trace on the Hilbert space associated with $M$.
The von Neumann entropy \cite{holevo2013quantum,wilde2017quantum} of the quantum system $M$ in the state $\rho$ is
\begin{equation}
S(M) = -\mathrm{Tr}\left[\rho\ln\rho\right]\;.
\end{equation}

A classical-quantum state of the classical random variable $X$ with values in $\mathbb{R}^n$ and of the quantum system $M$ is given by a probability density function $p(x)$ for $X$ and a collection of quantum states $\{\rho(x)\}_{x\in\mathbb{R}^n}$ on $M$, where for any $x\in\mathbb{R}^n$, $\rho(x)$ is the quantum state of $M$ conditioned on $X=x$.
In analogy with \eqref{eq:defccS}, we define the von Neumann entropy of $M$ conditioned on $X$ as \cite[Section III.A.3]{furrer2014position}, \cite{Murphy1990173}, \cite[Chapter 4.6-4.7]{takesaki2012theory}
\begin{equation}\label{eq:S(M|X)}
S(M|X) = \int_{\mathbb{R}^n}S(M|X=x)\,p(x)\,\mathrm{d}x\;,
\end{equation}
where for any $x\in\mathbb{R}^n$
\begin{equation}
S(M|X=x) = S(\rho(x))
\end{equation}
is the von Neumann entropy of $M$ conditioned on $X=x$.
When $M$ is a quantum system, conditioning on the values of $M$ does not have a well-defined meaning \cite{beigi2014dimension}, hence $S(X|M)$ cannot be defined as in \eqref{eq:defccS}.

When $M$ is a classical random variable, the chain rule for the entropy implies
\begin{equation}\label{eq:X|M}
S(X|M) = S(M|X) + S(X) - S(M)\;.
\end{equation}
Here
\begin{equation}
S(M) = -\sum_{m\in\mathcal{M}}q(m)\ln q(m)
\end{equation}
is the Shannon entropy \cite{cover2006elements} of $M$, where for any $m\in\mathcal{M}$, $q(m)$ is the probability that $M=m$, and $S(M|X)$ is defined by \eqref{eq:S(M|X)} with
\begin{equation}
S(M|X=x) = -\sum_{m\in\mathcal{M}}q(m|x)\ln q(m|x)\;,
\end{equation}
where $q(m|x)$ is the probability that $M=m$ conditioned on $X=x$.

All the entropies on the right-hand side of \eqref{eq:X|M} are still well-defined when $M$ is a quantum system, provided they are all finite.
We recall that
\begin{equation}
S(M) = S(\rho)\;,
\end{equation}
where
\begin{equation}
\rho = \int_{\mathbb{R}^n}\rho(x)\,p(x)\,\mathrm{d}x
\end{equation}
is the average state of $M$.
Therefore, we take \eqref{eq:X|M} as definition of the von Neumann entropy of $X$ conditioned on the quantum system $M$ \cite[Section III.A.3]{furrer2014position}, \cite{Murphy1990173}, \cite[Chapter 4.6-4.7]{takesaki2012theory}.

\subsection{Conditional independence}
Since conditioning on the values of a quantum system does not have a well-defined meaning, we cannot apply to a quantum conditioning the classical definition of conditional independence.
The random variables $X$ and $Y$ are conditionally independent given the random variable $M$ iff the mutual information between $X$ and $Y$ conditioned on $M$
\begin{equation}\label{eq:CMI}
I(X:Y|M) = S(X|M) + S(Y|M) - S(XY|M)
\end{equation}
vanishes \cite{cover2006elements}.
The right-hand side of \eqref{eq:CMI} is well-defined also when $M$ is a quantum system, hence we take \eqref{eq:CMI} as the definition of the mutual information conditioned on a quantum system \cite{holevo2013quantum,wilde2017quantum}, and we say that $X$ and $Y$ are conditionally independent given the quantum system $M$ iff
\begin{equation}
I(X:Y|M) = 0\;.
\end{equation}

A characterization theorem for the quantum states of a tripartite finite-dimensional quantum system with vanishing quantum conditional mutual information has been proven in \cite{hayden2004structure}.
In particular, the theorem can be applied to a tripartite finite-dimensional classical-classical-quantum system:
\begin{thm}\label{thm:independence}
Let $X$ and $Y$ be classical random variables with values in the finite sets $\mathcal{X}$ and $\mathcal{Y}$, respectively, and let $M$ be a finite-dimensional quantum system with Hilbert space $\mathcal{H}$.
Let us consider a classical-classical-quantum state on $XYM$ such that $X$ and $Y$ are conditionally independent given $M$, i.e.,
\begin{equation}
I(X:Y|M) = 0\;.
\end{equation}
Then, for any $x\in\mathcal{X}$ and any $y\in\mathcal{Y}$ the quantum state $\rho(x,y)$ of $M$ conditioned on $X=x$ and $Y=y$ has the form
\begin{equation}\label{eq:rhoind}
\rho(x,y) = \left.\bigoplus_{i=1}^N r(i)\,p(x|i)\,p(y|i)\,\rho_i^X(x)\otimes \rho_i^Y(y)\right/p(x,y)\;.
\end{equation}
Here $r(i)$ is a probability distribution on $\{1,\,\ldots,\,N\}$ and for any $i=1,\,\ldots,\,N$, $p(x|i)$ and $p(y|i)$ are probability distributions on $\mathcal{X}$ and $\mathcal{Y}$, respectively, such that for any $x\in\mathcal{X}$ and any $y\in\mathcal{Y}$
\begin{equation}i
\sum_{i=1}^N r(i)\,p(x|i)\,p(y|i) = p(x,y)\;,
\end{equation}
where $p(x,y)$ is the probability that $X=x$ and $Y=y$.
Moreover, for any $i=1,\,\ldots,\,N$, any $x\in\mathcal{X}$ and any $y\in\mathcal{Y}$, $\rho_i^X(x)$ and $\rho_i^Y(y)$ are quantum states on the Hilbert spaces $\mathcal{H}_i^X$ and $\mathcal{H}_i^Y$, respectively, such that
\begin{equation}
\mathcal{H} = \bigoplus_{i=1}^N\mathcal{H}_i^X\otimes \mathcal{H}_i^Y\;.
\end{equation}
\end{thm}
Formally Theorem \ref{thm:independence} has not been proven yet in the setup of this paper, where the classical random variables $X$ and $Y$ take continuous values.
In this case any classical-classical-quantum state of the form \eqref{eq:rhoind} satisfies $I(X:Y|M)=0$, but it is not known whether any state with $I(X:Y|M)=0$ has the form \eqref{eq:rhoind}.
\begin{rem}
A particularly simple and interesting class of classical-classical-quantum states with vanishing conditional mutual information is the case $N=1$ in Theorem \ref{thm:independence}, where the quantum state of $M$ conditioned on $X=x$ and $Y=y$ is
\begin{equation}
\rho(x,y) = \rho_X(x)\otimes\rho_Y(y)\;,
\end{equation}
and $\rho_X(x)$ and $\rho_Y(y)$ are quantum states on the Hilbert spaces $\mathcal{H}_X$ and $\mathcal{H}_Y$, respectively, such that
\begin{equation}
\mathcal{H} = \mathcal{H}_X\otimes\mathcal{H}_Y\;.
\end{equation}
\end{rem}

\section{The de Bruijn identity and the Stam inequality with quantum conditioning}\label{sec:Stam}
Our proof of the Entropy Power Inequality with quantum conditioning is based on the perturbation with the heat semigroup.
Let $X$ be a random variable with values in $\mathbb{R}^n$ and probability density function $p(x)$.
The probability density function $p(x,t)$ of $X$ evolved for time $t\ge0$ with the heat semigroup is the solution to the heat equation
\begin{equation}
\frac{\partial}{\partial t} p(x,t) = \nabla^2 p(x,t)\;,\qquad p(x,0)=p(x)\;.
\end{equation}
The fundamental property of the heat semigroup states that $p(x,t)$ coincides with the probability density function of the random variable $X + \sqrt{t}\,Z$, where $Z$ is a normal Gaussian random variable with values in $\mathbb{R}^n$ independent of $X$.
In the following, we will employ this representation of the heat semigroup.
The semigroup property implies
\begin{lem}[semigroup property]\label{lem:semi}
Let $X$ be a random variable with values in $\mathbb{R}^n$, and let $Z_1$, $Z_2$ and $Z$ be normal Gaussian random variables with values in $\mathbb{R}^n$ such that $X$, $Z_1$, $Z_2$ and $Z$ are all independent.
Then, for any $t_1,\,t_2\ge0$, $X+\sqrt{t_1}\,Z_1+\sqrt{t_2}\,Z_2$ has the same probability distribution as $X+\sqrt{t_1+t_2}\,Z$.
\end{lem}

\begin{defn}[Fisher information with quantum conditioning {\cite[Definition 8]{huber2018conditional}}]
Let $X$ be a random variable on $\mathbb{R}^n$ correlated with the quantum system $M$ such that $\|X\|^2$ has finite average and the function $x\mapsto \rho(x)$ is continuous in the trace norm, where for any $x\in\mathbb{R}^n$, $\rho(x)$ is the quantum state of $M$ conditioned on $X=x$.
The Fisher information of $X$ conditioned on $M$ is
\begin{equation}\label{eq:defJ}
J(X|M) = \lim_{t\to0^+} \frac{I\left(\left.X+\sqrt{t}\,Z:Z\right|M\right)}{t}\;,
\end{equation}
where $Z$ is a normal Gaussian random variable on $\mathbb{R}^n$ independent on both $X$ and $M$.
\end{defn}
\begin{rem}
The conditional mutual information in \eqref{eq:defJ} is computed on the classical-classical-quantum state of $X'=X+\sqrt{t}\,Z$, $Z$ and $M$.
This state is determined by the joint probability density of $X'$ and $Z$ and by the collection of quantum states $\{\rho'(x',z)\}_{x',\,z\in\mathbb{R}^n}$ of $M$ conditioned on $X'=x'$ and $Z=z$.
We recall that
\begin{equation}
\rho'(x',z) = \rho\left(x'-\sqrt{t}\,z\right)\;,
\end{equation}
where for any $x\in\mathbb{R}^n$, $\rho(x)$ is the quantum state of $M$ conditioned on $X=x$.
\end{rem}
\begin{rem}
The limit \eqref{eq:defJ} always exists, finite or infinite, from Lemma \ref{lem:conc} below.
\end{rem}
\begin{lem}[{\cite[Theorem 2]{huber2018conditional}}]\label{lem:conc}
The function
\begin{equation}
t\mapsto I\left(\left.X+\sqrt{t}\,Z:Z\right|M\right) = S\left(\left.X+\sqrt{t}\,Z\right|M\right) - S(X|M)
\end{equation}
is concave for any $t\ge0$.
\end{lem}
\begin{rem}
If the quantum system $M$ is trivial, the Fisher information with quantum conditioning coincides with the trace of the Fisher information matrix
\begin{equation}
J_{ij}(X) = \int_{\mathbb{R}^n}\frac{\partial\ln p(x)}{\partial x^i}\,\frac{\partial\ln p(x)}{\partial x^j}\,p(x)\,\mathrm{d}x\;,\qquad i,\,j=1,\,\ldots,\,n\;,
\end{equation}
where $p$ is the probability density function of $X$.
\end{rem}
The de Bruijn identity provides the link between the Fisher information and the growth rate of the entropy under the heat semigroup.
\begin{prop}[de Bruijn identity with quantum conditioning {\cite[Proposition 1]{huber2018conditional}}]\label{prop:dB}
Let $X$ be a random variable on $\mathbb{R}^n$ correlated with the quantum system $M$ such that $\|X\|^2$ has finite average and the function $x\mapsto \rho(x)$ is continuous in the trace norm, where for any $x\in\mathbb{R}^n$, $\rho(x)$ is the quantum state of $M$ conditioned on $X=x$.
Then,
\begin{equation}\label{eq:dB}
J(X|M) = \left.\frac{\mathrm{d}}{\mathrm{d}t}S\left(\left.X+\sqrt{t}\,Z\right|M\right)\right|_{t=0}\;,
\end{equation}
where $Z$ is a normal Gaussian random variable on $\mathbb{R}^n$ independent on both $X$ and $M$.
\end{prop}
\begin{rem}
If the quantum system $M$ is trivial, the de Bruijn identity with quantum conditioning \eqref{eq:dB} becomes the de Bruijn identity
\begin{equation}
J(X) = \left.\frac{\mathrm{d}}{\mathrm{d}t}S\left(X+\sqrt{t}\,Z\right)\right|_{t=0}
\end{equation}
exploited in the proof of the Entropy Power Inequality \eqref{eq:clEPI} of Ref. \cite{stam1959some}.
\end{rem}
\begin{cor}\label{cor:dB}
Let $X$ be a random variable on $\mathbb{R}^n$ correlated with the quantum system $M$ such that $\|X\|^2$ has finite average and the function $x\mapsto \rho(x)$ is continuous in the trace norm, where for any $x\in\mathbb{R}^n$, $\rho(x)$ is the quantum state of $M$ conditioned on $X=x$.
Let $Z$ be a normal Gaussian random variable on $\mathbb{R}^n$ independent on both $X$ and $M$.
Then, for any $t\ge0$
\begin{equation}
\frac{\mathrm{d}}{\mathrm{d}t}S\left(\left.X+\sqrt{t}\,Z\right|M\right) = J\left(\left.X+\sqrt{t}\,Z\right|M\right)\;,
\end{equation}
\begin{proof}
We have
\begin{align}
\frac{\mathrm{d}}{\mathrm{d}t}S\left(\left.X+\sqrt{t}\,Z\right|M\right) &= \left.\frac{\mathrm{d}}{\mathrm{d}s}S\left(\left.X+\sqrt{t+s}\,Z\right|M\right)\right|_{s=0}\nonumber\\
& = \left.\frac{\mathrm{d}}{\mathrm{d}s}S\left(\left.X+\sqrt{t}\,Z_1 + \sqrt{s}\,Z_2\right|M\right)\right|_{s=0}\nonumber\\
& = J\left(\left.X+\sqrt{t}\,Z_1\right|M\right)\;,
\end{align}
where $Z_1$ and $Z_2$ are independent normal Gaussian random variables uncorrelated from $X$, $Z$ and $M$, the second equality follows from Lemma \ref{lem:semi} and the last equality follows from Proposition \ref{prop:dB}.
\end{proof}
\end{cor}
\begin{thm}[Stam inequality with quantum conditioning]\label{thm:Stam}
Let $X$ and $Y$ be random variables with values in $\mathbb{R}^n$ correlated with the quantum system $M$ such that $\|X\|^2$ and $\|Y\|^2$ have finite average and the function $(x,y)\mapsto \rho(x,y)$ is continuous in the trace norm, where for any $x,\,y\in\mathbb{R}^{n}$, $\rho(x,y)$ is the quantum state of $M$ conditioned on $X=x$ and $Y=y$.
Let us assume that $X$ and $Y$ are conditionally independent given $M$, i.e.,
\begin{equation}\label{eq:hypind}
I(X:Y|M)=0\;.
\end{equation}
Then, for any $0\le\lambda\le1$ the following linear Stam inequality holds:
\begin{equation}\label{eq:linStam}
J(X+Y|M) \le \lambda^2 J(X|M) + \left(1-\lambda\right)^2 J(Y|M)\;.
\end{equation}
Choosing
\begin{equation}
\lambda = \frac{J(Y|M)}{J(X|M)+J(Y|M)}\;,
\end{equation}
\eqref{eq:linStam} becomes the Stam inequality
\begin{equation}\label{eq:Stam}
\frac{1}{J(X+Y|M)} \ge \frac{1}{J(X|M)} + \frac{1}{J(Y|M)}\;.
\end{equation}
\begin{proof}
We have
\begin{align}\label{eq:ineqI}
&I\left(\left.X+Y+\sqrt{t}\,Z:Z\right|M\right) \overset{(\text{a})}{\le} I\left(\left.X+\lambda\sqrt{t}\,Z\;,\;Y+\left(1-\lambda\right)\sqrt{t}\,Z:Z\right|M\right)\nonumber\\
&\overset{(\text{b})}{=} I\left(\left.X+\lambda\sqrt{t}\,Z:Z\right|M\right) + I\left(\left.Y+\left(1-\lambda\right)\sqrt{t}\,Z:Z\right|M\right) \nonumber\\
&+ I\left(\left.X+\lambda\sqrt{t}\,Z:Y+\left(1-\lambda\right)\sqrt{t}\,Z\right|MZ\right)\nonumber\\
& - I\left(\left.X+\lambda\sqrt{t}\,Z:Y+\left(1-\lambda\right)\sqrt{t}\,Z\right|M\right)\nonumber\\
&\overset{(\text{c})}{\le} I\left(\left.X+\lambda\sqrt{t}\,Z:Z\right|M\right) + I\left(\left.Y+\left(1-\lambda\right)\sqrt{t}\,Z:Z\right|M\right) \nonumber\\
&+ I\left(\left.X+\lambda\sqrt{t}\,Z:Y+\left(1-\lambda\right)\sqrt{t}\,Z\right|MZ\right)\nonumber\\
&\overset{(\text{d})}{=}  I\left(\left.X+\lambda\sqrt{t}\,Z:Z\right|M\right) + I\left(\left.Y+\left(1-\lambda\right)\sqrt{t}\,Z:Z\right|M\right)\;.
\end{align}
(a) and (b) follow from the data-processing inequality and the chain rule for the quantum conditional mutual information, respectively; (c) follows from the positivity of the quantum conditional mutual information; (d) follows from \eqref{eq:hypind} since
\begin{align}
I\left(\left.X+\lambda\sqrt{t}\,Z:Y+\left(1-\lambda\right)\sqrt{t}\,Z\right|MZ\right) &= I(X:Y|MZ)\nonumber\\
& = I(X:Y|M) = 0\;.
\end{align}
Since $Z$ is uncorrelated from $X$, $Y$ and $M$, we have
\begin{equation}
I(X+Y:Z|M) = I(X:Z|M) = I(Y:Z|M) = 0\;,
\end{equation}
hence both members of \eqref{eq:ineqI} vanish for $t=0$.
The claim \eqref{eq:linStam} then follows taking the derivative with respect to $t$ of both members of \eqref{eq:ineqI} at $t=0$.
\end{proof}
\end{thm}
\begin{rem}
If the quantum system $M$ is trivial, the Stam inequality with quantum conditioning \eqref{eq:Stam} becomes the Stam inequality
\begin{equation}
\frac{1}{J(X+Y)} \ge \frac{1}{J(X)} + \frac{1}{J(Y)}
\end{equation}
on which the proof of the Entropy Power Inequality \eqref{eq:clEPI} of Ref. \cite{stam1959some} is based.
\end{rem}
In the proof of the Entropy Power Inequality we also need the asymptotic scaling of the entropy under the heat semigroup evolution.
\begin{thm}[asymptotic scaling of entropy with quantum conditioning {\cite[Theorem 4]{huber2018conditional}}]\label{thm:asym}
Let $X$ be a random variable with values in $\mathbb{R}^n$ correlated with the quantum system $M$ such that $X$ and $M$ both have finite entropy.
Then,
\begin{equation}
\lim_{t\to\infty}\left(S\left(\left.X + \sqrt{t}\,Z\right|M\right) - n\,\frac{\ln t + 1}{2}\right) = 0\;,
\end{equation}
where $Z$ is a normal Gaussian random variable on $\mathbb{R}^n$ independent on both $X$ and $M$.
\end{thm}

\section{The Entropy Power Inequality with quantum conditioning}\label{sec:EPI}
\begin{thm}[Entropy Power Inequality with quantum conditioning]\label{thm:EPI}
Let $X$ and $Y$ be random variables with values in $\mathbb{R}^n$ correlated with the quantum system $M$ such that $\|X\|^2$ and $\|Y\|^2$ have finite average and the function $(x,y)\mapsto \rho(x,y)$ is continuous in the trace norm, where for any $x,\,y\in\mathbb{R}^{n}$, $\rho(x,y)$ is the quantum state of $M$ conditioned on $X=x$ and $Y=y$.
Let us assume that $X$ and $Y$ are conditionally independent given $M$, i.e.,
\begin{equation}\label{eq:hypind2}
I(X:Y|M)=0\;.
\end{equation}
Then, the following Entropy Power Inequality holds:
\begin{equation}\label{eq:EPI}
\exp\frac{2S(X+Y|M)}{n} \ge \exp\frac{2S(X|M)}{n} + \exp\frac{2S(Y|M)}{n}\;.
\end{equation}
\begin{proof}
We will prove the following linear Entropy Power Inequality for any $0\le\lambda\le1$:
\begin{equation}\label{eq:EPIlin}
S(X+Y|M) \ge \lambda\,S(X|M) + \left(1-\lambda\right)S(Y|M) - n\,\frac{\lambda\ln\lambda + \left(1-\lambda\right)\ln\left(1-\lambda\right)}{2}\;.
\end{equation}
The Entropy Power Inequality \eqref{eq:EPI} follows choosing
\begin{equation}
\lambda = \frac{\exp\frac{2S(X|M)}{n}}{\exp\frac{2S(X|M)}{n} + \exp\frac{2S(Y|M)}{n}}
\end{equation}
in \eqref{eq:EPIlin}.

Let $Z_1$, $Z_2$ and $Z$ be independent normal Gaussian random variables on $\mathbb{R}^n$ uncorrelated from both $X$ and $M$.
We define for any $t\ge0$
\begin{align}
\phi(t) &= S\left(\left.X + Y + \sqrt{t}\,Z\right|M\right)\nonumber\\
&\phantom{=} - \lambda\,S\left(\left.X + \sqrt{\lambda\,t}\,Z_1\right|M\right) - \left(1-\lambda\right)S\left(\left.Y + \sqrt{\left(1-\lambda\right)t}\,Z_2\right|M\right)\;.
\end{align}
We have from Corollary \ref{cor:dB}
\begin{align}\label{eq:phi'}
&\phi'(t) = J\left(\left.X + Y + \sqrt{t}\,Z\right|M\right)\nonumber\\
& - \lambda^2\,J\left(\left.X + \sqrt{\lambda\,t}\,Z_1\right|M\right) - \left(1-\lambda\right)^2J\left(\left.Y + \sqrt{\left(1-\lambda\right)t}\,Z_2\right|M\right)\;.
\end{align}
From \eqref{eq:hypind2}, $X + \sqrt{\lambda\,t}\,Z_1$ and $Y + \sqrt{\left(1-\lambda\right)t}\,Z_2$ are conditionally independent given $M$ since from the data-processing inequality for the quantum conditional mutual information we have
\begin{equation}
I\left(\left.X + \sqrt{\lambda\,t}\,Z_1:Y + \sqrt{\left(1-\lambda\right)t}\,Z_2\right|M\right) \le I(X:Y|M) = 0\;.
\end{equation}
Theorem \ref{thm:Stam} with the replacements $X\mapsto X + \sqrt{\lambda\,t}\,Z_1$ and $Y\mapsto Y + \sqrt{\left(1-\lambda\right)t}\,Z_2$ implies
\begin{align}\label{eq:Stam2}
&J\left(\left.X + \sqrt{\lambda\,t}\,Z_1 + Y + \sqrt{\left(1-\lambda\right)t}\,Z_2\right|M\right)\nonumber\\
& \le \lambda^2\,J\left(\left.X + \sqrt{\lambda\,t}\,Z_1\right|M\right) + \left(1-\lambda\right)^2 J\left(\left.Y + \sqrt{\left(1-\lambda\right)t}\,Z_2\right|M\right)\;.
\end{align}
We have from Lemma \ref{lem:semi}
\begin{equation}
J\left(\left.X + \sqrt{\lambda\,t}\,Z_1 + Y + \sqrt{\left(1-\lambda\right)t}\,Z_2\right|M\right) = J\left(\left.X + Y + \sqrt{t}\,Z\right|M\right)\;,
\end{equation}
hence \eqref{eq:phi'} and \eqref{eq:Stam2} imply
\begin{equation}\label{eq:ineqphi}
\phi'(t)\le0\;.
\end{equation}
We have from Lemma \ref{lem:semi} again
\begin{equation}
S\left(\left.X + \sqrt{\lambda\,t}\,Z_1 + Y + \sqrt{\left(1-\lambda\right)t}\,Z_2\right|M\right) = S\left(\left.X + Y + \sqrt{t}\,Z\right|M\right)\;.
\end{equation}
Then, from Lemma \ref{lem:conc}, $\phi(t)$ is a linear combination of concave functions, hence from \eqref{eq:ineqphi}
\begin{equation}
\phi(t) = \phi(0) + \int_0^t\phi'(s)\,\mathrm{d}s \le \phi(0)\;.
\end{equation}
Hence, we have from Theorem \ref{thm:asym}
\begin{equation}
\phi(0) \ge \limsup_{t\to\infty}\phi(t) = -n\,\frac{\lambda\ln\lambda + \left(1-\lambda\right)\ln\left(1-\lambda\right)}{2}\;,
\end{equation}
and the claim \eqref{eq:EPIlin} follows.
\end{proof}
\end{thm}

\section{Conclusions}\label{sec:concl}
We have proven the Entropy Power Inequality with quantum conditioning (Theorem \ref{thm:EPI}).
This inequality complements the Entropy Power Inequality with classical conditioning \cite{wang2017distributed}, the conditional Entropy Power Inequality for bosonic quantum systems \cite{de2018conditional} and the conditional Entropy Power Inequality for quantum additive-noise channels.
The Entropy Power Inequality with classical conditioning was necessary in the field of distributed source coding to prove the converse theorems for the quadratic Gaussian CEO problem with two terminals \cite{oohama2005rate,prabhakaran2004rate,wang2017distributed} and for the Gaussian multi-terminal source coding problem with two sources \cite{gamal2011network,wang2017distributed}.
In the same way, the Entropy Power Inequality with quantum conditioning proven in this paper can be useful to prove the converse theorems for the quantum versions of the same problems, where the encoders can exploit a shared entangled quantum state.
The hypothesis of conditional independence is not a limitation to the applicability of the Entropy Power Inequality with quantum conditioning, since the same hypothesis is necessary for the Entropy Power Inequality with classical conditioning and fulfilled in all its applications.

\section*{Acknowledgements}
GdP acknowledges financial support from the European Research Council (ERC Grant Agreements Nos. 337603 and 321029), the Danish Council for Independent Research (Sapere Aude), VILLUM FONDEN via the QMATH Centre of Excellence (Grant No. 10059), and the Marie Sk\l odowska-Curie Action GENIUS (Grant No. 792557).

\includegraphics[width=0.05\textwidth]{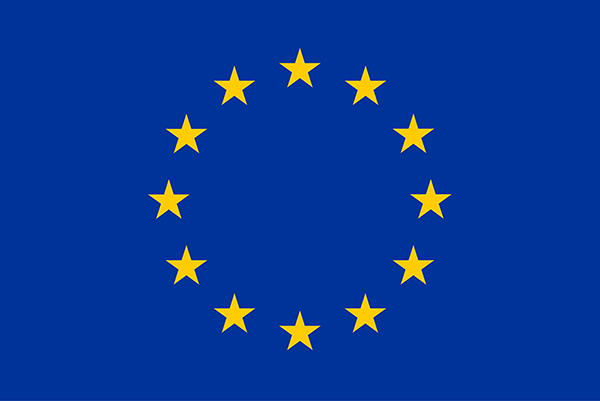}
This project has received funding from the European Union's Horizon 2020 research and innovation programme under the Marie Sk\l odowska-Curie grant agreement No. 792557.

\bibliography{biblio}
\bibliographystyle{unsrt}
\end{document}